\documentstyle[aps,prl,epsf]{revtex}
\begin{document}
\draft

\title{Large-N solutions of the Heisenberg and Hubbard-Heisenberg models
on the anisotropic triangular lattice:  application to
Cs$_2$CuCl$_4$ and to the layered organic 
superconductors $\kappa$-(BEDT-TTF)$_2$X}

\author{C. H. Chung and J. B. Marston}

\address{Department of Physics, Brown University, Providence, RI 
02912-1843, USA}

\author{Ross H. McKenzie}

\address{Department of Physics, University of Queensland,
Brisbane 4072, Australia}

\date{\today}

\maketitle

\begin{abstract}
We solve the Sp(N) Heisenberg and SU(N) Hubbard-Heisenberg models on
the anisotropic triangular lattice in the large-N limit. These two models
may describe respectively the magnetic and electronic properties of the family
of layered organic materials $\kappa$-(BEDT-TTF)$_2$X.
The Heisenberg model is also relevant to the frustrated
antiferromagnet, Cs$_2$CuCl$_4$.  We find rich
phase diagrams for each model.  The Sp(N) antiferromagnet is shown to have 
five different phases as a function of the size of the spin and the degree of
anisotropy of the triangular lattice.
The effects of fluctuations at finite-N are also discussed.
For parameters relevant to Cs$_2$CuCl$_4$ the ground state either
exhibits incommensurate spin order, or is in a quantum disordered
phase with deconfined spin-1/2 excitations and topological order.
The SU(N) Hubbard-Heisenberg model exhibits an insulating
dimer phase, an insulating box phase, a semi-metallic staggered flux phase (SFP), 
and a metallic uniform phase. The uniform and SFP phases
exhibit a pseudogap.  A metal-insulator transition occurs at
intermediate values of the interaction strength.

\end{abstract}


\section{Introduction}
\label{sec:intro}

The family of layered organic superconductors $\kappa$-(BEDT-TTF)$_2$X
has attracted much experimental and theoretical
interest\cite{ishiguro,wosnitza}. There are many similarities to the
high-$T_{c}$ cuprates, including unconventional metallic properties
and competition between antiferromagnetism and superconductivity\cite{mckenzie}.  
The materials have a rich phase diagram 
as a function of pressure and temperature.  At low
pressures and temperatures there is an insulating antiferromagnetic ordered phase; 
as the temperature is increased a transition occurs to an insulating paramagnetic state. 
A first-order metal-insulator transition separates the 
paramagnetic insulating phase from a metallic phase; it is induced by
increasing the pressure\cite{ito,komatsu}.  The metallic phase exhibits
various temperature dependences which are different from that of conventional 
metals. For example, measurements of the magnetic susceptibility
and NMR Knight shift are consistent with a weak
pseudogap in the density of states\cite{desoto,mayaffre}.

The main purpose of this paper is to attempt to describe the magnetic ordering,
the metal-insulator transition, and the unconventional metallic
properties of these materials with two simplified models.  We
model the magnetic ordering in the insulating phase with the quantum
Heisenberg antiferromagnet (HAF).  We model the metallic phase 
as well as the metal-insulator
transition with a hybrid Hubbard-Heisenberg model.  To substitute for 
the lack of a small expansion parameter in either model, we enlarge the symmetry
group from the physical SU(2) $\cong$ Sp(1) spin symmetry 
to Sp(N) (symplectic group) for the Heisenberg model\cite{subir91,sachdev0} 
and to SU(N) for the Hubbard-Heisenberg model\cite{affleck,marston}. 
We then solve these models in the large-N limit and treat 
$1/N$ as our systematic expansion parameter.  In section \ref{sec:models} 
we briefly summarize the
physical SU(2) Heisenberg and Hubbard-Heisenberg model on the anisotropic 
triangular lattice.  In section \ref{sec:spn} we review the large-N theory of the Sp(N) 
quantum Heisenberg model.  Based on the large-N solution of this model, 
we present the magnetic phase diagram in the parameter space of quantum
fluctuation $n_b/N$ (where $n_b$ is the number of
bosons in the Schwinger boson representation of
the spin) and the magnetic frustration $J_2/J_1$. We
discuss the effects of finite-N fluctuations on
the Sp(N) magnetic phases with short-range order (SRO).
We also discuss how our results are relevant to understanding 
recent neutron scattering experiments on Cs$_2$CuCl$_4$.
In section \ref{sec:sun} we review the large-N theory
of the SU(N) Hubbard-Heisenberg model. We present the phase diagram 
based on the large-N solution in the parameter space of the dimensionless ratio 
of the nearest-neighbor exchange to the hopping constant $J_1/t_1$ and the
dimensionless anisotropy ratio $J_2/(J_1+J_2)$.  Away from the two nested limits
$J_1 = 0$ and $J_2 = 0$ we find a metal-insulator transition which occurs at 
finite critical value of 
$J_1/t_1$.  We also find that the density of states in
the metallic state is suppressed at low temperatures, in qualitative 
agreement with the unconventional metallic properties seen in experiments.   
We conclude in section \ref{sec:discuss} with a brief review our results.

\section{Heisenberg and Hubbard-Heisenberg Models on the Anisotropic
Triangular Lattice}
\label{sec:models}
  
\begin{figure}
\epsfxsize = 8cm
\hspace{5.5cm}
\epsfbox{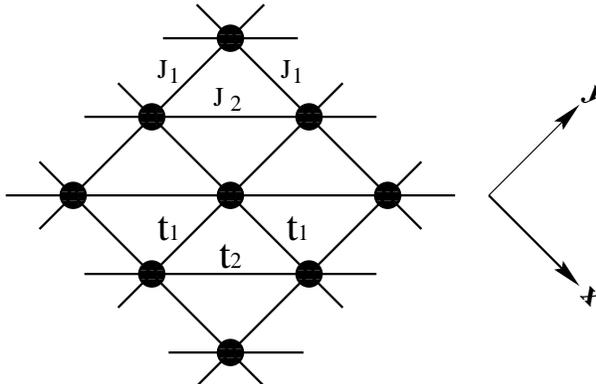}
\caption{The anisotropic triangular lattice with two types of bonds. 
Note that this can also be viewed as a square lattice with an 
additional next-nearest neighbor interaction along only one of the
two diagonals.}
\label{lattice}
\end{figure}

Based on a wide range of experimental results and
quantum chemistry calculations of the Coulomb repulsion
between two electrons on the BEDT-TTF molecules it was argued
in reference \onlinecite{mckenzie}
that the $\kappa$-(BEDT-TTF)$_2$X family
are strongly correlated electron 
systems which can be described by a half-filled Hubbard model 
on the anisotropic triangular lattice.  The Hubbard 
Hamiltonian is: 
\begin{eqnarray}
H &=& 
-t_1 \sum_{<{\bf{ij}}>}[c^{\dagger \sigma}_{\bf{i}}
c_{{\bf{j}}\sigma}+ H.c.] - t_2\sum_{<<{\bf{ij}}>>}
[c^{\dagger \sigma}_{\bf{i}} c_{{\bf{j}} \sigma}+ H.c.]
\nonumber \\
&+&
\frac{U}{2} \sum_{\bf{i}}(c^{\dagger \sigma}_{\bf{i}}c_{{\bf{i}} \sigma}-1)^2.
\end{eqnarray}
Here $c_{{\bf{i}}\sigma}$ is the electron destruction operator on site
${\bf{i}}$ and there is an implicit sum over pairs of raised and lowered
spin indices $\sigma = \uparrow, \downarrow$. 
Matrix element $t_1$ is the nearest-neighbor hopping amplitude, 
and $t_2$ is the next-nearest-neighbor hopping along only one of the 
two diagonals of the square lattice as shown in figure \ref{lattice}.
The sum $<{\bf{ij}}>$ runs over pairs of nearest-neighbor sites and 
$<<{\bf{ij}}>>$ runs for next-nearest-neighbors. 

Physical insight can be attained by considering the Hubbard model for
different values of the ratio $U/t$.
In the limit of large $U/t$ the Hubbard model at half-filling 
is insulating and the spin degrees of freedom are
described by a spin-1/2 Heisenberg antiferromagnet\cite{auerbach}:
\begin{eqnarray}
H &=& 
J_1\sum_{<{\bf{ij}}>} \vec{S}_{\bf{i}}\cdot \vec{S}_{\bf{j}} + 
J_2\sum_{<<{\bf{ij}}>>} \vec{S}_{\bf{i}}\cdot \vec{S}_{\bf{j}}, 
\end{eqnarray}
where $\vec{S}_{\bf{i}}$ is the spin operator on site $\bf{i}$, and the exchange
couplings $J_1 = 4t_1^2/U$ and $J_2 = 4t_2^2/U$. Competition between 
$J_1$ and $J_2$ leads to magnetic frustration.  Using parameters
from quantum chemistry 
calculations\cite{fortu,okuno,castet} it was estimated
in reference \onlinecite{mckenzie} that $J_2/J_1 \sim 0.3$ to $1$
for the $\kappa$-(BEDT-TTF)$_2$X family.  Hence,
magnetic frustration is important.
The frustrated Heisenberg Hamiltonian interpolates between the 
square-lattice ($J_2 = 0$) and the linear chain ($J_1 = 0$).  Much is
known about these two limiting cases.  
Additional insight\cite{trumper,merino} comes by considering
different values of the ratio $J_2/J_1$. At 
$J_2 = 0$, the square lattice limit, there is long-range N\'eel
order with a magnetic moment of approximately $0.6 \mu_B$; see 
reference \cite{auerbach}. If 
$J_2$ is small but nonzero, the magnetic moment will be reduced by 
magnetic frustration.  At $J_2/J_1$ around $0.5$, quantum fluctuations 
combined with frustration should destroy the N\'eel ordered state.
As $J_2/J_1$ is further increased, the system may exhibit spiral long 
range order\cite{merino}.  At $J_1 = J_2$, the lattice is equivalent to 
the isotropic triangular lattice.  Anderson suggested in 1973
that the ground state could be a spin liquid 
without long range order\cite{anderson}.  
However, subsequent numerical work indicated that 
there is long-range-order with ordering vector
$\vec{q}=(2\pi/3, 2\pi/3)$\cite{sorella}.  Finally, at $J_1 = 0$
the model reduces to decoupled Heisenberg spin-1/2 chains which cannot 
sustain long range spin order\cite{auerbach}
as per the quantum Mermin-Wagner theorem. For 
$J_1$ small but non-zero, the system consists of spin-1/2 chains weakly
coupled in a zigzag fashion. The case of two such weakly coupled 
zigzag spin chains was studied by Okamoto and Nomura\cite{okamoto} and
by White and Affleck\cite{white} who showed that there is a spin gap
in the spectrum $\Delta \sim e^{-{\rm const} \times J_2/J_1}$, and the
ground state exhibits dimerization and incommensurate spiral correlations. 
Although our system consists of an infinite number of weakly coupled spin 
chains instead of two chains, we find similar behavior.  

As $U/t$ decreases, charge fluctuations from electron hopping become 
significant.  Competition between hopping and 
Coulomb repulsion leads to a transition from the insulator to a metal. 
We use the hybrid Hubbard-Heisenberg Hamiltonian\cite{affleck,marston} with independent 
parameters $t$, $J$, and $U$ (where in general $J \neq 4t^2/U$) to 
describe some aspects of the transition: 
\begin{eqnarray}
H &=& 
\sum_{<{\bf{ij}}>}[-t_1 (c^{\dagger \sigma}_{\bf{i}}
c_{{\bf{j}}\sigma} + H.c.) + J_1 (\vec{S}_{\bf{i}} \cdot \vec{S}_{\bf{j}}
- \frac{1}{4} n_{\bf i} n_{\bf j})]
\nonumber \\
&+& 
\sum_{<<{\bf{ij}}>>}[-t_2 (c^{\dagger \sigma}_{\bf{i}} c_{{\bf{j}}\sigma}+ H.c.) +
J_2 (\vec{S}_{\bf{i}} \cdot \vec{S}_{\bf{j}} - \frac{1}{4} n_{\bf i} n_{\bf j})]
\nonumber \\
&+&
\frac{U}{2}\sum_{\bf{i}}(c^{\dagger \sigma}_{\bf{i}}c_{{\bf{i}}\sigma}-1)^2\ .
\label{HSU2}
\end{eqnarray}  
Here $n_{\bf i} \equiv c_{\bf i}^{\dagger \sigma} c_{\bf {i} \sigma}$
is the number of fermions at site $\bf{i}$.  The Hamiltonian reduces to the 
Hubbard model when $J_1=J_2=0$ and the Heisenberg model for $t_1, t_2 \to 0$. 
In the square lattice limit and at half-filling, 
perfect nesting drives the system to an antiferromagnetic insulator at 
arbitrarily small value of the interactions. As diagonal hopping $t_2$
is turned on, however, nesting of the Fermi surface is no longer perfect and 
the metal-insulator transition (MIT) occurs at non-zero critical interaction strength.

\section{Sp(N) Heisenberg Model} 
\label{sec:spn}

We first focus on the spin degree of freedom, appropriate to the insulating
phase of the layered organic materials, by solving the Heisenberg model in 
a large-N limit.  Our choice of the large-N generalization of the physical
SU(2) spin-$1/2$ problem is dictated by the desire to find an exactly 
solvable model which has both long-range ordered (LRO) and short-range ordered 
(SRO) phases.  This leads us to symmetric (bosonic)
SU(N) or Sp(N) generalizations.  The former can only
be applied to bipartite lattices.
The latter approach has been applied to the antiferromagnetic Heisenberg 
model on the square lattice
with first-, second-, and third-neighbor 
coupling (the $J_1 - J_2 - J_3$ model)\cite{subir91},
the isotropic triangular lattice\cite{subir92}, and
the kagom\'e lattice\cite{subir92}.
As the anisotropic triangular lattice is not
bipartite, we must choose the Sp(N) generalization\cite{subir91}.

\subsection{Brief Review of the Approach} 
To ascertain the likely phase diagram of the 
frustrated Heisenberg antiferromagnet, we consider the Sp(N) symplectic group 
generalization of the physical 
SU(2)$\cong$ Sp(1) antiferromagnet\cite{subir91,subir92}.
The model can be exactly solved in the $N \rightarrow \infty$ limit.  
Both LRO and SRO
can arise if we use the symmetric (bosonic) representations of Sp(N).  We begin
with the bosonic description of the SU(2) HAF, where it can be shown that apart from an 
additive constant, the Hamiltonian may be written in terms of spin singlet bond operators:
\begin{equation}
H = -\frac{1}{2}\sum_{\bf{ij}} J_{\bf{ij}}
(\varepsilon_{\alpha\beta}
b_{\bf{i}}^{\dagger \alpha} b_{\bf{j}}^{\dagger \beta}) 
(\varepsilon^{\gamma \delta} b_{\bf{i} \gamma} b_{\bf{j} \delta}) 
\label{Hsp1}
\end{equation}
where we have used the bosonic representation for spin operator 
\begin{eqnarray}
\vec{S}_{\bf i} = \frac{1}{2} b_{\bf i}^{\dagger \alpha} 
\vec{\sigma}_{\alpha}^{\beta} b_{{\bf i} \beta}\ ,
\end{eqnarray} 
where $\alpha = \uparrow, \downarrow$ labels the two possible spins of each boson.
The antisymmetric tensor, $\varepsilon_{\alpha\beta}$ is as usual defined to be:
\begin{eqnarray}
\varepsilon = \left(
\begin{array}{cc}
0&1\\
-1&0
\end{array} \right)\ .
\end{eqnarray}
We enforce the constraint $n_b = b_{\bf{i}}^{\dagger \alpha} b_{\bf{i} \alpha} = 2S$ 
to fix the number of bosons, and hence the total spin, on each site.  
The SU(2) spin singlet 
bond creation operator $\varepsilon_{\alpha\beta} b_{\bf{i}}^{\dagger \alpha}
b_{\bf{j}}^{\dagger \beta}$ may now be generalized to the Sp(N)-invariant 
form ${\mathcal J}_{\alpha\beta} b_{\bf{i}}^{\dagger \alpha} b_{\bf{j}}^{\dagger \beta}$.
Global Sp(N) rotations may be implemented with $2N \times 2N$ unitary matrices $U$:
\begin{eqnarray}
b &\to& U b 
\nonumber \\
U^{\dagger} {\mathcal J} U &=& {\mathcal J}.
\end{eqnarray}
Here 
\begin{equation}
{\mathcal J} = 
\left( \begin{array}{cccccc}
 & 1 & & & & \\
-1 &  & & & & \\
 & &  & 1 & & \\
 & & -1 & & & \\
 & &  & & \ddots & \\
 & & & & & \ddots
\end{array} \right)
\end{equation}
generalizes the $\epsilon$ tensor to $N > 1$, and 
$b_{\bf{i}\alpha}$ with $\alpha = 1 \ldots 2N$ is the Sp(N) boson
destruction operator.  The Hamiltonian of the Sp(N) HAF may then 
be written:
\begin{equation}
H = -\sum_{\bf{ij}} \frac{J_{\bf{i j}}}{2N} 
\left({\mathcal J}_{\alpha \beta}
b_{\bf{i}}^{\dagger \alpha} b_{\bf{j}}^{\dagger \beta} \right)
\left( {\mathcal J}^{\gamma \delta} b_{\bf{i} \gamma} b_{\bf{j} \delta}
\right)
\label{Hspn}
\end{equation}
where again the Greek indices run over $1 \ldots 2N$, and 
the constraint $b_{\bf{i}}^{\dagger \alpha} b_{\bf{i} \alpha} = n_b$ is 
imposed at every site of the lattice.  We have rescaled $J_{\bf{i j}}/2 \rightarrow
J_{\bf{i j}}/ 2N$ to make the Hamiltonian of $O(N)$.  For fixed $J_1/J_2$ we have a
two-parameter ($n_b$, $N$) family of models with the ratio $\kappa \equiv n_b / N$
determining the strength of the quantum fluctuations.  In the physical Sp(1) limit,
$\kappa = 1$ corresponds to spin-1/2.  At large $\kappa$ (equivalent to the large-spin
limit of the physical SU(2) model) quantum effects are small and the ground states 
break global Sp(N) spin-rotational symmetry.  LRO then corresponds to Bose 
condensation which we quantify by defining
\begin{equation}
b_{\bf{i} m \sigma} \equiv \left ( \begin{array}{c}
\sqrt{N} x_{{\bf i} \sigma}\\
\tilde{b}_{{\bf i} \tilde{m} \sigma}
\end{array} \right )\ .
\label{bx}
\end{equation}
Here the spin-quantization axis has been fixed by introducing the paired-index 
notation $\alpha \equiv (m, \sigma)$ with $m = 1 \ldots N$, $\sigma = \uparrow,\downarrow$, 
and $\tilde{m}=2 \ldots N$.  The c-number spinors $x_\sigma$, when non-zero, quantify the
Bose condensate fraction, and are given by 
$\langle b_{{\bf i} m \sigma} \rangle = \sqrt{N} \delta^1_m x_{\bf{i} \sigma}$.
For sufficiently small $\kappa$, however, quantum fluctuations 
overwhelm the tendency to order and there can be only magnetic SRO.  

Upon inserting equation \ref{bx} into equation \ref{Hspn}, decoupling the 
quartic boson terms within the corresponding functional integral by introducing
complex-valued Hubbard-Stratonovich fields $Q_{\bf{ij}}$ directed along the lattice links, 
enforcing the constraint on the number of bosons on each site 
with Lagrange multipliers $\lambda_{\bf{i}}$, and finally integrating out the $b$-fields, 
we obtain an effective action proportional to $N$. Thus, as $N \to \infty$, 
the effective action may be replaced by its saddle-point value.  
At the saddle-point, the $Q_{\bf{ij}}$, $\lambda_{\bf{i}}$, and  
$x_{\bf{i} \sigma}$ fields are expected to be time-independent, so
the effective action can be put into correspondence with a suitable 
mean-field Hamiltonian.  The Hamiltonian may be diagonalized by a 
Bogoliubov transformation, yielding a total ground state
energy $E_{MF}$ given by\cite{subir92}
\begin{eqnarray}
\frac{E_{MF}[Q, \lambda]}{N} 
&=& \sum_{\bf{i>j}}\left(
\frac{J_{\bf{ij}}}{2}{|Q_{\bf{ij}}|}^{2} - 
\frac{J_{\bf{ij}}}{2}Q_{\bf{ij}}\varepsilon^{\sigma\sigma^{\prime}}
x_{\bf{i} \sigma} x_{\bf{j} \sigma^{\prime}} + H.c.\right) 
\nonumber \\
&-& 
\sum_{\bf{i}}\lambda_{\bf{i}}\left(1+\frac{n_b}{N}
- |x_{\bf{i} \sigma}|^{2} \right) + \sum_{\bf k} \omega_{\bf k}
\label{EMF}
\end{eqnarray}
where $\omega_{\bf k}$ are eigenenergies obtained
from diagonalizing the mean-field Hamiltonian.  Finding the ground state of the
Sp(N) HAF now reduces to the problem of minimizing $E_{MF}$ with 
respect to the variables $Q_{{\bf{ij}}}$ and $x_{\bf{i} \sigma}$, subject to 
the Lagrange-multiplier constraints 
\begin{equation}
\frac{\partial E_{MF}[Q, \lambda]}{\partial \lambda_{\bf i}} = 0\ .
\label{lagrange}
\end{equation}

It is essential to note that the action possesses local U(1) gauge symmetry
under local phase rotations by angle $\theta_i(\tau)$: 
\begin{eqnarray}
b_{{\bf i}\alpha} \to b_{{\bf i}\alpha}~ e^{-i \theta_{\bf{i}}(\tau)}
\nonumber \\
Q_{\bf{ij}} \to Q_{\bf{ij}}~ e^{i \theta_{\bf{i}}(\tau) + i\theta_{\bf{j}}(\tau)}
\nonumber \\
\lambda_{\bf{i}} \to \lambda_{\bf{i}} + \frac{\partial \theta_{\bf{i}}}
{\partial \tau}\ .
\label{U(1)}
\end{eqnarray}
This symmetry reminds us that the representation of spin operators
in terms of the underlying bosons is redundant as the phase of each boson fields 
can be shifted by an arbitrary amount without affecting the spin degree of freedom.   
Two gauge invariant quantities of particular importance for our classification of 
the phases are $|Q_{\bf {ij}}|$, and $\sum_{\bf {i}} \lambda_{\bf i}$.
 
Extrema of $E_{MF}$ are found numerically with the simplex-annealing 
method\cite{NRC}.  We work with a lattice of $40 \times 40 $ sites and check that this is
sufficiently large to accurately represent the thermodynamic limit.  Constraint
equation \ref{lagrange} is tricky, however, as $E_{MF}[\lambda_{\bf i}]$ 
is neither a minimum, nor
a maximum with respect to the $\lambda_{\bf i}$ directions at the saddle point.  
The problem is solved by decomposing $\lambda_{\bf{i}}$ into its mean value $\bar{\lambda}$ 
and the deviations from the mean, $\delta \lambda_{\bf i} \equiv \lambda_{\bf i} - \bar{\lambda}$.  
As $\bar{\lambda}$ is gauge-invariant, we solve the constraint equation \ref{lagrange} separately
for it by applying Newton's method.  The resulting 
$E_{MF}[\bar{\lambda}, \delta \lambda_{\bf i}]$ can then be maximized in the 
remaining $\delta \lambda_{\bf i}$ directions.  Individual 
values of $\delta \lambda_{\bf{i}}$ are in general non-zero, but by definition 
it must be the case that $\sum_{\bf i} \delta\lambda_{\bf{i}} = 0$.  

\subsection{Phase Diagram of the Sp(N) Heisenberg antiferromagnet}
\begin{figure}
\epsfxsize = 6.5cm
\hspace{5.5cm}
\epsfbox{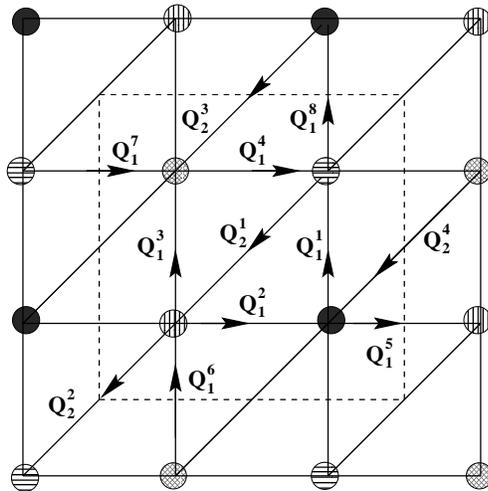}
\caption{The $2 \times 2$-site unit cell for the Sp(N) Heisenberg antiferromagnet
on the anisotropic triangular lattice. 
The complex-valued, directed, $Q_1^1$ to $Q_1^8$ fields live on the links 
of the square lattice, while $Q_2^1$ to $Q_2^4$ live on the diagonal links.  
Arrows denote the orientation of the $Q_{\bf ij}$ fields.} 
\label{4sitespn}
\end{figure}

\begin{figure}
\epsfxsize = 5.5cm
\hspace{5.5cm}
\epsfbox{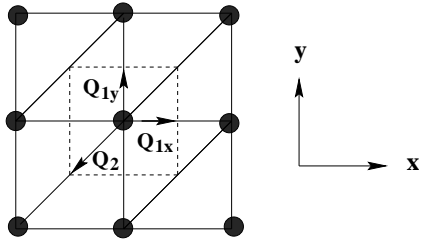}
\caption{The one-site unit cell for the Sp(N) Heisenberg antiferromagnet
on the anisotropic triangular lattice.} 
\label{1sitespn}
\end{figure}

To make progress in actually solving the model, we now make the assumption that
spontaneous symmetry breaking, if it occurs, does not lead to a 
unit cell larger than 4 sites.  Our choice of unit cell is shown in 
figure \ref{4sitespn}.
The $2 \times 2$ unit cell requires 12 different complex-valued Q-fields 
(8 on the square links, and 4 along the diagonal links) and 4 different 
$\lambda$ fields and $x$ spinors at each of the four sites.  However, we have
checked that at every saddle-point in the SRO region of the phase diagram
($x_{{\bf i} \sigma} = 0$)
each of the 8 Q-fields on the horizontal and vertical links take the same value.
Likewise, the 4 Q-fields on the diagonals are all equal, as well as 
all 4 $\lambda$-fields.  Thus the $2 \times 2$-site unit cell can be
reduced to only a single site per unit cell as shown 
in figure \ref{1sitespn}.  We expect this to also hold in the LRO
phases, in accord with previous work on the Sp(N) model\cite{subir92}.

The zero-temperature phase diagram is a function of two variables: $J_2/J_1$ 
and $\kappa$.  The various saddle points can be classified in several ways. 
Both SRO and LRO phases may be characterized in terms of an ordering wavevector 
$\vec{q}$ via the relation $\vec{q} = 2 \vec{k}_{min}$, where $\vec{k}_{min}$ is the wavevector 
at which the bosonic spinon energy spectrum has a minimum.  The spin structure 
factor $S(\bf{q})$ peaks at this wavevector\cite{subir92}.  LRO is signaled by 
non-zero spin condensate $x_{\bf{i} \sigma}$, which we assume occurs at only 
one wavevector, $\vec{k}_{min}$; that is, 
$x_{\bf{j} \uparrow} = x \exp(i \vec{k}_{min} \cdot \vec{r}_{\bf j})$ 
and $x_{\bf{j} \downarrow} = -i x \exp(-i \vec{k}_{min} \cdot \vec{r}_{\bf j})$. 
The phases may be further classified\cite{subir91} according to the particular 
value of $\vec{q}$.  There can be commensurate collinear ordering tendencies 
where the spins rotate with a period that
is commensurate with the underlining lattice.  Alternatively, there may be 
incommensurate coplanar ordering tendencies where the spins rotate in a two dimensional spin 
space with a period that is not commensurate with the lattice.

\begin{figure}
\epsfxsize = 12.0cm
\hspace{2.0cm}
\epsfbox{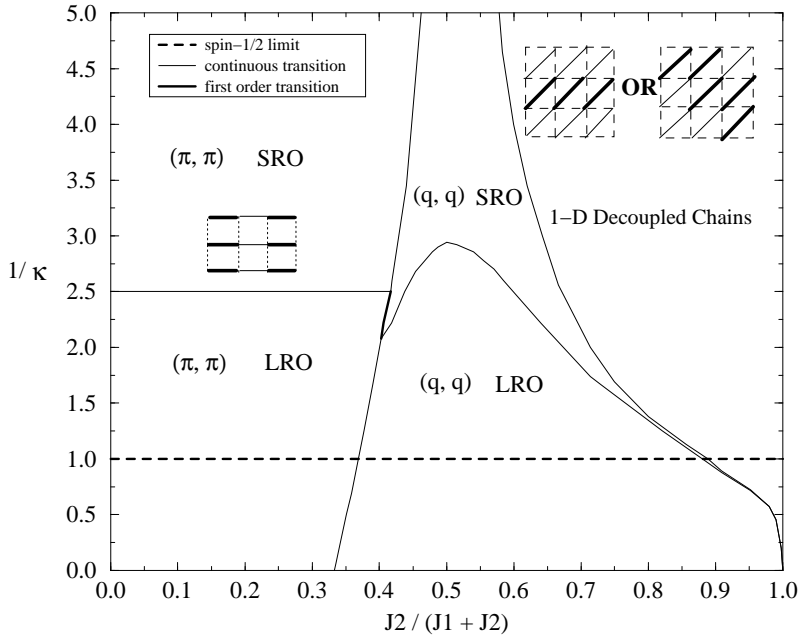}
\caption{Zero-temperature phase diagram of the Sp(N) Heisenberg antiferromagnet at 
large-N.  The strength of the quantum fluctuations is set by the parameter $1/\kappa$ 
(vertical axis).  Dimerization patterns induced by finite-N fluctuations are shown in
the insets.}
\label{spnphase}
\end{figure}

The phase diagram is shown in figure \ref{spnphase}.
Note that the general shape of the phase diagram
is qualitatively consistent with the
finding of spin wave calculations\cite{merino,trumper}
that quantum fluctuations are largest for
$J_2 / J_1 \sim 0.5$ and large $J_2/J_1$.  For large enough 
values of $\kappa$, the ground state has magnetic LRO.  
As the magnetic LRO phases break Sp(N) 
symmetry, there are gapless Goldstone spin-wave modes.  As a check
on the calculation, we note that in the 
$\kappa \rightarrow \infty$ limit there is a transition
between the N\'eel ordered and incommensurate $(q, q)$ LRO phase at
$J_2 / J_1 = 0.5$ in agreement with the classical large-spin limit.
At smaller values of $\kappa$ there are quantum disordered phases that 
preserve global Sp(N) symmetry.  In the $N \rightarrow \infty$ limit these
are rather featureless spin liquids with gapped, free, spin-1/2 bosonic 
excitations (spinons).  Finite-N fluctuations, however, induce qualitative 
changes to the commensurate SRO phases (see below).  In the limiting case
of the nearest-neighbor square lattice, $J_2 = 0$, we reproduce the previously
obtained result\cite{subir91} that N\'eel order arises for $\kappa > 0.4$.  In the 
opposite limit of decoupled one dimensional spin chains, $J_1 = 0$, 
the ground state is in a disordered phase at all values of $\kappa$. 
There are five phases in all, three commensurate and two incommensurate, as
detailed in the following two subsections. 

\subsubsection{Commensurate phases} 
At small to moderate $J_2/J_1$ there are two phases 
with $Q_{1x} = Q_{1y} \neq 0$, and $Q_2 = 0.$  
The eigenspectrum $\omega_{\bf k}$ has its minimum at 
${\bf k}= \pm(\pi/2, \pi/2)$, with the implication 
that the spin-spin correlation function peaks at $\vec{q} = (\pi, \pi)$.  
N\'eel LRO with $x_{\bf{i} \sigma} \neq 0$ appears when $\kappa$ is
sufficiently large.  The boundary between LRO and SRO phases is 
independent of $J_2/J_1$ except at one end of the boundary, 
but this is expected to be an artifact of the large-N limit\cite{subir92}. 
Finite-N corrections should bend this horizontal phase boundary. 

At large $J_2/J_1$ and small $\kappa$ the ground state is
a disordered state characterized by $Q_{1x} = Q_{1y} = 0$ and $Q_2 \neq 0$. 
The chains decouple from one another exactly in the large-N limit, 
but at any finite-N the chains will be coupled by the fluctuations about
the saddle point.  Also, $x_{\bf{i} \sigma} = 0$ as it must, by the Mermin-Wagner theorem. 
The Sp(N) solution does not properly describe the 
physics of completely decoupled spin-$1/2$ chains.  
All spin excitations are gapped, and 
there is dimerization at finite-N (see below).  This 
behavior is in marked contrast to the gapless, undimerized ground state
of the physical SU(2) spin-$1/2$ nearest-neighbor Heisenberg chain\cite{auerbach}.  

\subsubsection{Incommensurate phases} 
At intermediate values of $J_2/J_1$, there are two incommensurate phases with 
$Q_{1x} = Q_{1y} \neq 0$, and $Q_2 \neq 0$.  The ordering wavevector 
$\vec{q} = (q, q)$ with $q$ varying continuously from $\pi$ to 
$\pi/2$, a sign of helical spin order in a given plane. The 
inverse $\kappa_c^{-1}$ of the critical $\kappa$ separating LRO from SRO
peaks near the isotropic triangular point $J_2/J_1 = 1$, where it agrees
in value to that reported in a prior study of the triangular lattice\cite{subir92}, 
and then decreases with a long tail as $J_2/J_1$ increases.  As the one-dimensional 
limit of decoupled chains is approached, ${\kappa}_c \rightarrow \infty$.
Again this accords with the Mermin-Wagner theorem. 

All the phase transitions are continuous except for the transitions between the
$(\pi, \pi)$ LRO phase and the $(q, q)$ SRO phase which is first order. 
Fluctuations at finite-N, however, modify the mean-field
results\cite{subir91}.  The modifications are only quantitative 
for the LRO phases, and for the incommensurate SRO phase.  
In particular, instantons in the U(1) gauge field have little effect in 
the incommensurate SRO phase which is characterized by 
non-zero $Q_2$.  The $Q_2$ fields carry charge $\pm 2$, so 
when they acquire a non-zero expectation value, this is equivalent to 
the condensation of a charge-2 Higgs field.  Fradkin 
and Shenker\cite{fradkin} showed some time ago that a Higgs condensate 
in 2+1 dimensions quenches the confining U(1) gauge force.  
Singly-charged spinons are therefore deconfined, instead the 
instantons which carry magnetic flux are confined, 
and no dimerization is induced\cite{subir91}. 
The relevant non-linear sigma model which describes 
the transition from an ordered incommensurate
phase to a quantum disordered phase has been studied 
by Chubukov, Sachdev, and Senthil\cite{chubukov}.

In the case of the commensurate $(\pi, \pi)$ SRO and the decoupled
chain phases, instantons {\it do} alter the states qualitatively.  The Berry's phases 
associated with the instantons lead to columnar spin-Peierls 
order\cite{subir90,subir91} (equivalent to dimer order)
as indicated in figure \ref{spnphase}.  The dimerization pattern induced
in the decoupled chain phase is similar to that found by White and 
Affleck\cite{white} for a pair of chains with zigzag coupling.  
Furthermore, spinon excitations are confined into pairs by the U(1) gauge force.  

Note that the deconfined spinons in the $(q, q)$ SRO phase
are qualitatively different than the spinons found
in the limit of completely decoupled chains, $J_1 = 0$.  
The $(q, q)$ SRO phase exhibits true 2-dimensional fractionalization,
in contrast to the decoupled one-dimensional chains. The spinons are massive,
again in contrast to those found in one-dimensional half-odd-integer spin chains.  
The transition from the dimerized chain phase at small $J_1/J_2$, which has confined spinons, to
the deconfined incommensurate phase at larger $J_1/J_2$ is described by a 2+1 dimensional 
Z$_2$ gauge theory\cite{subir00}.  In fact the $(q, q)$ SRO phase is similar to 
a resonating valence bond (RVB) state recently found on the isotropic triangular lattice 
quantum dimer model\cite{moessner}.  The phase has ``topological'' order; 
consequently when the lattice is placed on a torus (that is, when periodic boundary conditions
are imposed), the ground state becomes four-fold degenerate 
in the thermodynamic limit\cite{topological,cms}. 

\subsection{Physical Spin-$1/2$ Limit}
\begin{figure}
\epsfxsize = 12.0cm
\hspace{2.0cm}
\epsfbox{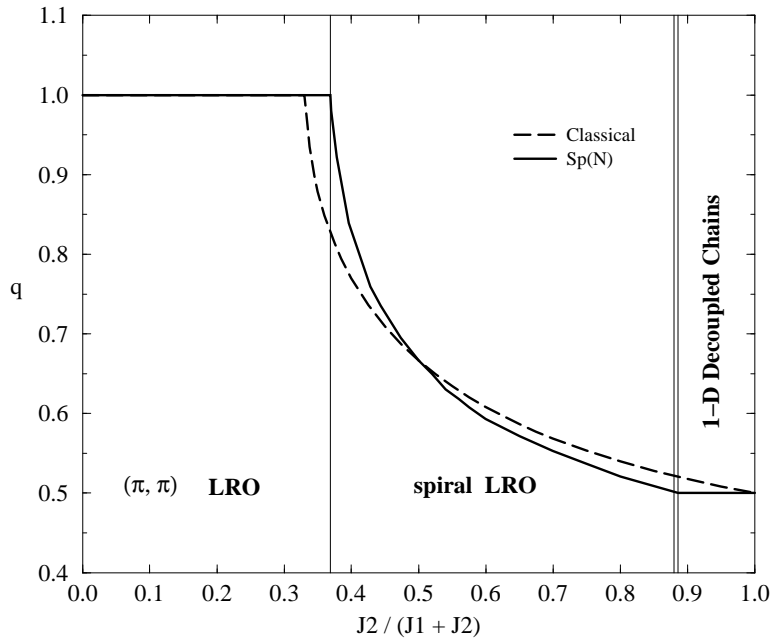}
\caption{Ordering wavevector $q$ (in units of $\pi/a$) of the large-N Sp(N) Heisenberg 
antiferromagnet at $\kappa = 1$, which corresponds to
spin-1/2 in the physical Sp(1) limit. 
The solid line is the classical ordering wavevector\protect\cite{merino}.}
\label{spinhalf}
\end{figure}

It is interesting to examine in more detail the physical 
spin-$1/2$ limit corresponding to $\kappa = 1$.  
In figure \ref{spinhalf} we plot the 
ordering wavevector $q$ as a function of the ratio $J_2/(J_1 + J_2)$.
Note that quantum fluctuations cause 
(i) the N\'eel phase to be stable for larger values
of $J_2/J_1$ than classically, and
(ii) the wave vector associated with the
incommensurate phases deviates from
the classical value.  Similar behavior was also
found in studies based on a series expansion\cite{zhang}, and
slave bosons including fluctuations about the saddle point\cite{ceccatto}.  

\begin{figure}

\epsfxsize = 9cm
\hspace{3.5cm}
\epsfbox{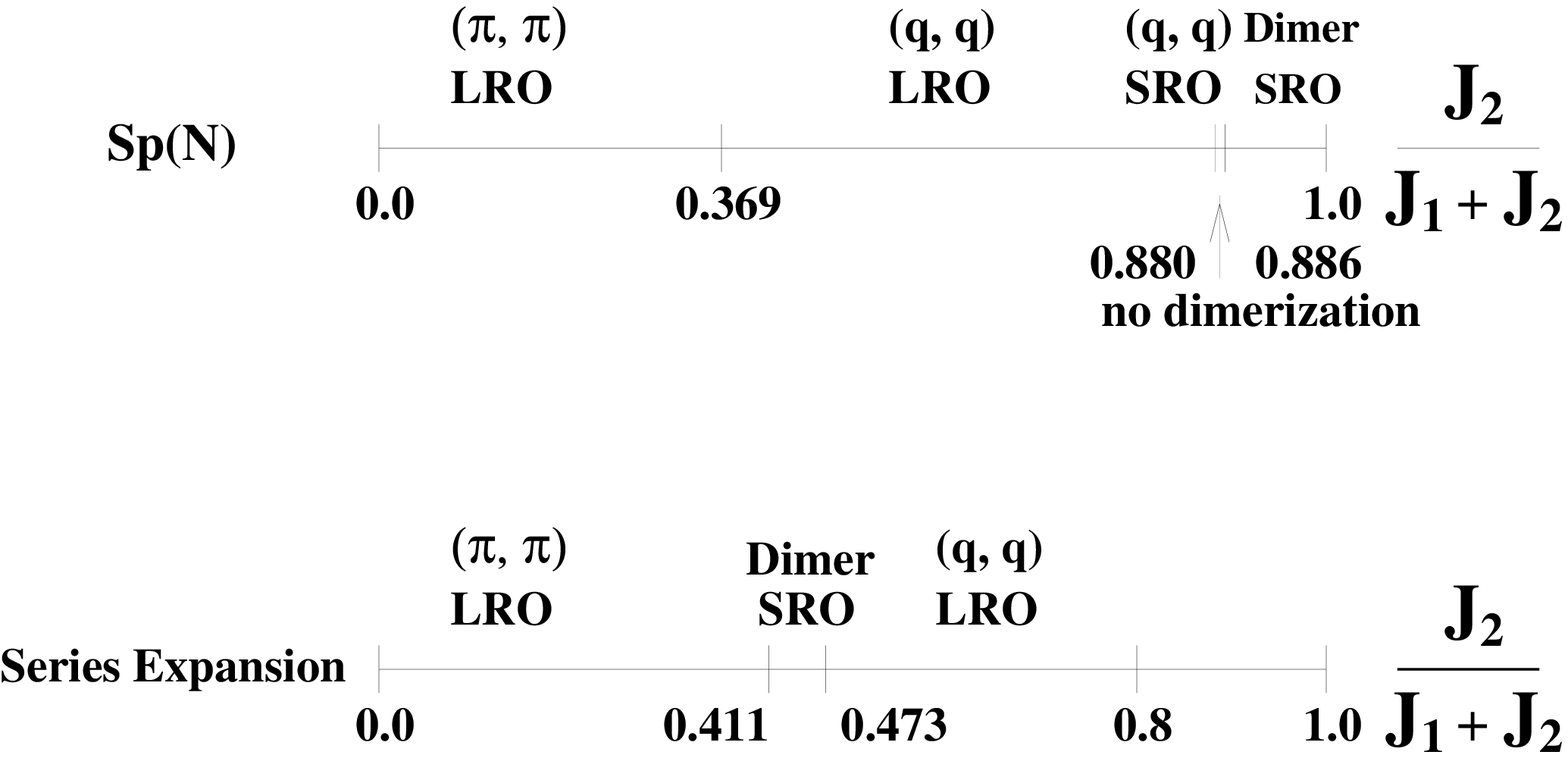}
\caption{Comparison of Sp(N) phase diagram at $\kappa = 1$ with the results of a
series-expansion\protect\cite{zhang}.}
\label{spincom}
\end{figure}

Commensurate $q = \pi$ N\'eel order persists up to $J_2/(J_1+J_2) = 0.369$, and
this is followed by the incommensurate LRO phase for 
$0.369 < J_2/(J_1+J_2) < 0.880$.  Then a tiny sliver of the incommensurate SRO phase 
arises for
$0.880 < J_2/(J_1+J_2) < 0.886$.  Finally there is the decoupled chain phase for 
$0.886 < J_2/(J_1+J_2) \leq 1$.  A strikingly similar phase diagram has been obtained by 
the series expansion method\cite{zhang}.  A comparison between the Sp(N)
and series expansion results is shown in figure \ref{spincom}.
Both methods suggest that there exists a narrow SRO region between N\'eel and 
incommensurate ordered phases, though in the Sp(N) case the region does not 
extend all the way up to $\kappa = 1$.  Possibly finite-$N$ corrections to
this large-N result could change the phase diagram quantitatively such that
the $(q, q)$ SRO phase persists up to $\kappa = 1$ in agreement with the 
series expansion results.  Another difference is that the Sp(N) result 
shows no dimerization in this narrow SRO region (due to the 
above-mentioned Higgs mechanism), while the series expansion indicates 
possible dimer order.

Similar results have also been obtained in a weak-coupling renormalization group 
(RG) treatment of the Hubbard model on the anisotropic triangular 
lattice\cite{tsai}.  For the special case of a pure square lattice (with
next-nearest-neighbor hopping $t_2 = 0$) 
at half-filling, the antiferromagnetic (AF) couplings 
diverge much faster during the RG transformations than couplings in the 
BCS sector, indicating a tendency towards magnetic LRO.  As 
$t_2$ is turned on, BCS and AF instabilities begin to compete.  
For sufficiently large $t_2$, a crossover occurs to a $d_{x^2-y^2}$ BCS instability, 
suggesting that the system is now in a magnetic SRO state.  
Further increasing $t_2$ to reach the isotropic
triangular lattice ($t_1 = t_2$) there are indications that long-range 
AF order re-enters.  Finally, for $t_2 \gg t_1$ LRO tendencies are again
eliminated, this time by the strong one-dimensional fluctuations.  It is remarkable
that the same sequence of ordering and disordering tendencies 
-- LRO to SRO to LRO to SRO -- occurs in all three approximate solutions.   

\subsection{Application To Materials}
The region of the phase diagram intermediate between the square lattice and 
the isotropic triangular lattice is relevant to the insulating phase of 
the organic $\kappa$-(BEDT-TTF)$_2$X materials.  The expected range in 
the spin exchange interaction\cite{mckenzie} is $J_2/J_1 \sim 0.3$ to $1$. 
Depending on the precise ratio $J_2/J_1$, our phase diagram indicates
that these materials could be in the N\'eel ordered phase, 
the $(q, q)$ LRO phase, or possibly the paramagnetic $(q, q)$ SRO phase
(see above). In fact antiferromagnetic ordering with a magnetic moment of $0.4$ to 
$1 \times \mu_B$ per dimer has been seen in the splitting of
proton NMR lines in the $\kappa$-(BEDT-TTF)$_2$Cu[N(CN)$_2$]Cl 
compound\cite{kanoda}.  It is conceivable that a quantum phase transition 
from the N\'eel ordered phase to the paramagnetic  
$(q, q)$ SRO phase or to the $(q, q)$ LRO phase
can be induced by changing the anion X. 

Coldea {\it et al.}\cite{coldea} have recently performed a
comprehensive neutron scattering study of Cs$_2$CuCl$_4$.
They suggest that this material is described by the spin-1/2 version
of our model with $J_2/J_1 = 2.5$ and $J_2 = 0.37$ meV.
The measured incommensurability with respect to N\'eel order, $\pi - q$, 
is reduced below the classical value by a factor $0.47$,
consistent with series results\cite{zhang}    
(our Sp(N) solution shows a smaller reduction).
Maps of the excitation spectra show that the
observed dispersion is renormalized upward in energy 
by a factor of $1.67$, which can be compared to
theoretical values of $1.18$ for the square lattice and
$1.57$ for decoupled chains.
Furthermore, the dynamical structure factor $S(\vec{q}, \omega)$ does
not exhibit well defined peaks corresponding
to well defined spin-1 magnon excitations.
Instead there is a continuum of excitations similar
to those expected and seen in completely decoupled spin-$1/2$ Heisenberg chains.  
In the case of a chain these excitations can be interpreted 
as deconfined gapless spin-1/2 spinons.

\begin{figure}
\epsfxsize = 9cm
\hspace{3.5cm}
\epsfbox{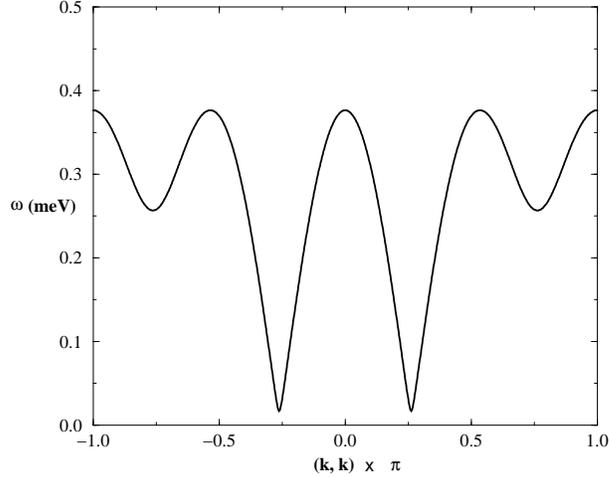}
\caption{Spinon dispersion along the $(1, 1)$ direction in the deconfined $(q, q)$ SRO
phase for  $\kappa = 0.56$, $J1 = 0.148$ meV, and $J2 = 0.37$ meV; 
thus, $J2/J1 = 2.5$.  The lowest energy excitations occur
at an incommensurate wavevector of $k \approx 0.26 \times \pi$ where there 
is a non-zero energy gap of approximately $0.02$ meV.} 
\label{spinonSRO}
\end{figure}

In the relevant parameter regime, $J_2/J_1$ = 2.5, the large-N Sp(N) phase diagram 
predicts that the ground state is spin-ordered with an incommensurate 
wavevector $(q, q)$.  However, as noted above, finite-N corrections could 
move the phase boundaries so that the physical spin-1/2 limit is in fact
described by the $(q, q)$ SRO phase.  In this phase there is 
a non-zero gap to the lowest lying excitations [which occur at wave vector 
$(k_{min}, k_{min}) = (q/2, q/2)$] rather than gapless excitations.  
At $\kappa = 0.56$ and $J_2/J_1 = 2.5$, in the large-N limit, the system is in the 
$(q, q)$ SRO phase of figure \ref{spnphase} and 
the spinon gap is approximately $0.05 J_2$.  This is much smaller than the 
resolution of the experiment [see Figure 2(a) in reference \cite{coldea}],
in which excitation energies have only been measured
down to about 0.5 $J_2$.  The corresponding spinon dispersion 
is shown in figure \ref{spinonSRO}.   Again we stress that 
the spinons in the incommensurate SRO phase are qualitatively
different than those which arise in a one-dimensional chain.  

\begin{figure}
\epsfxsize = 9cm
\hspace{3.5cm}
\epsfbox{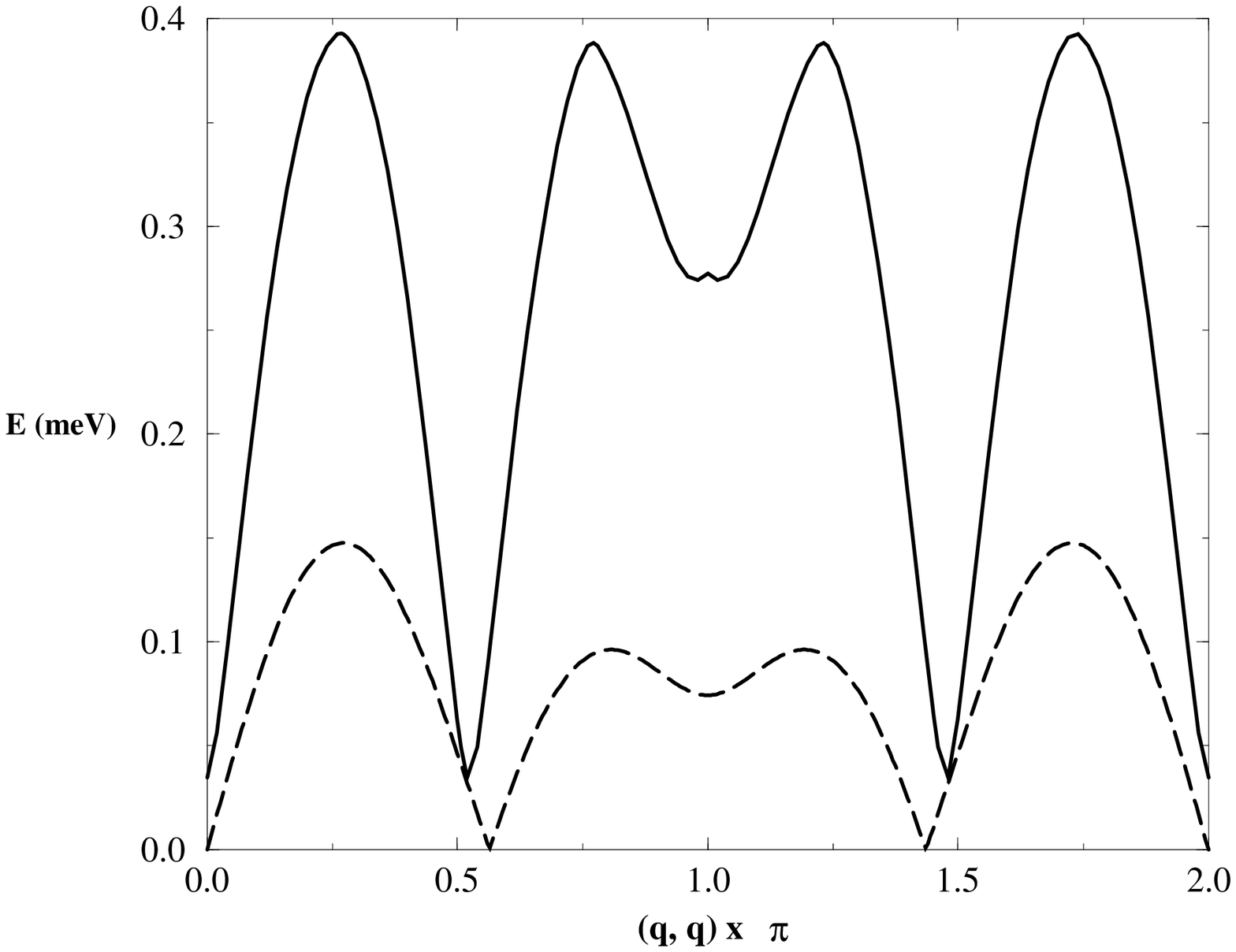}
\caption{Minimum energy of two spinons in the $(q, q)$ SRO phase (solid line) 
compared to that of classical spinwaves (dashed line)
for the same parameters as in figure \ref{spinonSRO}.  
The dispersion, in the case of the Sp(N) theory, is obtained from equation \ref{twospinon}.
The classical spinwave dispersion is obtained from 
reference \protect\cite{merino} where it was scaled to spin-$1/2$.  These 
energies have been multiplied here by a factor of $\kappa^2 \approx 0.31$ 
to account for the reduced magnitude of the spin.  
The large upward renormalization in the Sp(N) excitation energies
compared to the classical energies is due to quantum fluctuations.}
\label{spinwaveSRO}
\end{figure}

In the deconfined $(q, q)$ SRO phase, there are no true spinwave excitations, as 
spin rotational symmetry is unbroken.  Nevertheless, when the gap to create a spinon
is small, the spinwave description remains useful.  For example, in neutron scattering
experiments, spinons are created in pairs, as each spinon carries spin-$1/2$.  So a spinwave
may be viewed as an excitation composed of two spinons, though of course this spinwave does
not exhibit the sharp spectral features of a true Goldstone mode.  At large-N the spinons 
do not interact; finite-N fluctuations will lead to corrections in the combined energy and to
a finite lifetime. 
The minimum energy of such a spinwave of momentum $\vec{q}$ is given, in the large-N limit, by:
\begin{equation}
E_{\vec{q}} = {\rm Min}\{\omega(\vec{q}/2 + \vec{p}/2) + \omega(\vec{q}/2 - \vec{p}/2)\}
\label{twospinon}
\end{equation} 
where the minimization is with respect to all possible relative momenta $\vec{p}$.
The resulting spinwave dispersion is plotted in figure \ref{spinwaveSRO} alongside
the classical result, scaled to the same value of the spin.  
The Sp(N) calculation shows a rather large upward 
renormalization in the energy scale compared to the classical calculation; 
this is a result of the quantum fluctuations which are retained in the large-N limit.  

\begin{figure}
\epsfxsize = 9cm
\hspace{3.5cm}
\epsfbox{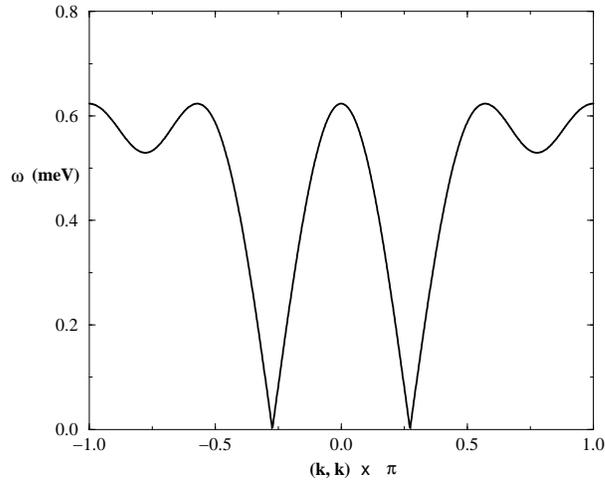}
\caption{Spinon dispersion along the $(1, 1)$ direction in the incommensurate $(q, q)$ LRO
phase.  As in figure \ref{spinonSRO}, $J1 = 0.148$ meV and $J2 = 0.37$ meV, but now
$\kappa = 1$.  Gapless excitations occur at an incommensurate wavevector of 
$k \approx 0.27 \times \pi$.} 
\label{spinon}
\end{figure}

\begin{figure}
\epsfxsize = 9cm
\hspace{3.5cm}
\epsfbox{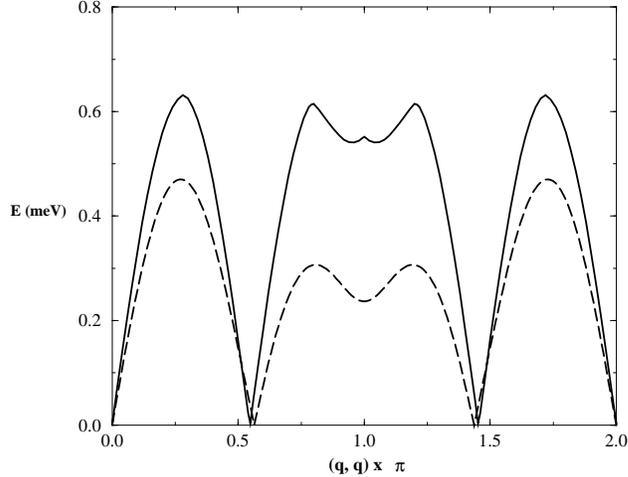}
\caption{The same as figure \ref{spinwaveSRO} except in the 
incommensurate $(q, q)$ LRO phase with the same parameters as in figure \ref{spinon}:
$J1 = 0.148$ meV and $J2 = 0.37$ meV, and $\kappa = 1$.}
\label{spinwave}
\end{figure}

In the incommensurate $(q, q)$ LRO phase ($\kappa = 1$ and $N \rightarrow \infty$) 
the spinon spectrum has gapless excitations, as shown in figure \ref{spinon}.  
Apart from the absence of the small gap, the spinwave dispersion in the ordered phase 
is similar to the minimum energy spectrum in the disordered phase. Again there is 
an upward renormalization of the energy scale as shown in figure \ref{spinwave},
though the ratio is relatively smaller than in the more 
quantum $\kappa = 0.56$ case.  The size of the renormalization is in good agreement with 
that seen in the Cs$_2$CuCl$_4$ experiment\cite{coldea}.  It is important to note
that finite-N gauge fluctuations bind spinons in the LRO phase into true spinwave excitations, 
with corresponding sharp spectral features.  In contrast, as noted above, 
spectral weight is smeared out in the SRO phase.  A large spread of spectral weight 
is seen in the neutron scattering experiments\cite{coldea}.
But as the incommensurate SRO and LRO states 
are separated by a continuous phase transition (see figure \ref{spnphase}), 
in the vicinity of the phase boundary it is difficult to distinguish the two types
of excitation spectra.  The spin moment in the LRO phase is small there, as is the gap
in the SRO phase.  Further experiments on Cs$_2$CuCl$_4$ may be needed to determine which of the 
two phases is actually realized in the material.  

\section{SU(N) Hubbard-Heisenberg Model}
\label{sec:sun}

We now turn our attention to the charge degrees of freedom by studying the
hybrid Hubbard-Heisenberg model.  This model should provide a reasonable  
description of the layered organic materials because the Hubbard interaction
is comparable in size to the hopping matrix elements, $U \approx t$.  Thus
the stringent no-double-occupancy constraint of the popular $t-J$
model should be relaxed.  In the large-N limit it is also better to work with  
the hybrid Hubbard-Heisenberg model than with the pure Hubbard model
because the crucial spin-exchange
processes are retained in the large-N limit\cite{marston}.  In the Hubbard model these
are only of order $1/N$ and therefore vanish in the mean-field description. 

As there are now both charge and spin degrees of freedom, it is no longer possible
to employ purely bosonic variables, in contrast to the previous section.  Instead we use
antisymmetric representations of the group SU(N) as the large-N generalization of 
the physical SU(2) system.  As shown below, this generalization 
precludes the possibility of describing magnetic LRO or superconducting phases, at least
in the large-N limit.  However, as the same representation is placed on each lattice site,
this large-N generalization works equally well for bipartite and non-bipartite lattices.  
It has been applied to the spin-1/2 Heisenberg antiferromagnet on the kagom\'e 
lattice\cite{zeng}.   

\subsection{Brief Review of the Approach}
The Hubbard-Heisenberg Hamiltonian on the anisotropic 
triangular lattice is specified by equation \ref{HSU2}.  The generalized
SU(N) version is obtained\cite{marston} by simply letting the spin index 
$\sigma$ in equation \ref{HSU2} run from 1 to N (where N is even):
\begin{eqnarray}
H &=& 
\sum_{<{\bf{ij}}>}[-t_1 (c^{\dagger\sigma}_{\bf{i}} c_{{\bf{j}}\sigma}+ H.c.) -
\frac{J_1}{N}(c^{\dagger\alpha}_{\bf{i}}
c_{{\bf{j}}\alpha}c^{\dagger\beta}_{\bf{j}}c_{{\bf{i}}\beta}+
\frac{1}{2} n_{\bf{i}}n_{\bf{j}})]
\nonumber \\
&+& 
\sum_{<<{\bf{ij}}>>}[-t_2 (c^{\dagger\sigma}_{\bf{i}}
c_{{\bf{j}}\sigma} + H.c.) - \frac{J_2}{N}(c^{\dagger\alpha}_{\bf{i}}
c_{{\bf{j}}\alpha}c^{\dagger\beta}_{\bf{j}}c_{{\bf{i}}\beta}+
\frac{1}{2} n_{\bf{i}}n_{\bf{j}})]
\nonumber \\
&+&
\frac{U}{N}\sum_{\bf{i}}(c^{\dagger\sigma}_{\bf{i}}c_{{\bf{i}}\sigma} - N/2)^2\ 
\label{HSUN}
\end{eqnarray}
where all spin indices are summed over.
Here we have also rescaled the interaction strengths
$J_i/2 \rightarrow J_i/N$
and $U/2 \rightarrow U/N$ to make each of the terms in the Hamiltonian of 
order N.  At half-filling, the only case we consider here, a further 
simplification occurs as the term $J_{\bf{ij}}n_{\bf{i}}n_{\bf{j}}$ is 
simply a constant.  There is no possibility of phase separation into 
hole-rich and hole-poor regions, nor can stripes form\cite{matthias}, as the system
is at half-filling.  We drop this constant term in the following analysis. 

We could also include a biquadratic spin-spin interaction of the form
$\tilde{J} (\vec{S}_{\bf i} \cdot \vec{S}_{\bf j})^2$ with $\tilde{J} > 0$.  
In the physical SU(2) limit
this term does nothing except renormalize the strength of the usual bilinear 
spin-spin interaction.  But for $N > 2$ it suppresses 
dimerization\cite{marston}, as 
the concentration of singlet correlations on isolated bonds is particularly
costly when the biquadratic term is included.  Thus there exists a family
of large-N theories parameterized by the dimensionless ratio $\tilde{J}/J$, 
each of which has the same physical SU(2) limit.  In this paper, however, 
we set $\tilde{J} = 0$ as we find that the phase most likely to 
describe the metallic regime of the organic superconductors is a uniform
phase with no dimerization which is stable even at $\tilde{J} = 0$.   

After passing to the functional-integral formulation in terms of Grassman fields,   
the quartic interactions are decoupled by a Hubbard-Stratonovich 
transformation which introduces real-valued auxiliary fields $\phi$
on each site and complex-valued $\chi$ fields directed along each bond:
\begin{eqnarray}
\phi_{\bf{i}} = 
{\it i} \frac{U}{N} (c^{* \sigma}_{\bf{i}} c_{{\bf{i}}\sigma} - N/2),\quad
\chi_{\bf{i}\bf{j}} = 
\frac{J_{\bf{i}\bf{j}}}{N}c^{* \sigma}_{\bf{i}} c_{{\bf{j}}\sigma}\ .
\label{phichi}
\end{eqnarray}
Clearly $\langle \phi_{\bf{i}} \rangle$ is proportional to the local 
charge density relative to half-filling (corresponding to $N/2$ fermions at
each site) and $\chi_{\bf{i}\bf{j}}$ may be viewed as an effective
hopping amplitude for the fermions between site $\bf{i}$ and $\bf{j}$. The
effective action in terms of these auxiliary fields, which may be viewed
as order parameters, can now be obtained by integrating out the fermions.
We note that, unlike the bosonic formulation of the pure antiferromagnet, 
here there is no possibility of magnetic LRO as the order parameters
$\chi$ and $\phi$ are
global SU(N) invariants, and of course there is no possibility of Bose
condensates in the fermionic antisymmetric representation of SU(N).  Superconductivity
likewise is not possible in the large-N limit because the order parameters are
invariant under global U(1) charge symmetry rotations.    
In the Heisenberg limit $t\to 0$ the action is also invariant under  
local U(1) gauge transformations as long as the $\chi$- and $\phi$-fields 
transform as gauge fields: $\chi_{\bf ij}(\tau) \to e^{[i\theta_{\bf i}(\tau) - 
\theta_{\bf j}(\tau)]} \chi_{\bf ij}(\tau)$, $\phi_{\bf i}(\tau ) \to \phi_{\bf i}(\tau ) - 
d\theta_{\bf i}(\tau)/d\tau$.  For the more general case of 
Hubbard-Heisenberg model, this local U(1) gauge symmetry breaks to only 
global U(1) gauge symmetry reflecting the conservation of total charge. 

Since $S_{eff}$ has an overall factor of N, the saddle-point 
approximation is exact at $N \to \infty$.  We expect $\phi$ and $\chi$ to be
time-independent at the saddle point, so $S_{eff}$ can be written in terms
of the free energy of fermions moving in a static order-parameter background: 
\begin{eqnarray}
S_{eff}[\phi, \chi] &=& \beta F[\phi, \chi; \mu]\ ,
\nonumber \\ 
{{F[\phi, \chi; \mu]}\over{N}} &=&
\sum_{<\bf{ij}>} \frac{|\chi_{\bf ij}|^2}{J_1}
+ \sum_{<<\bf{ij}>>}\frac{|\chi_{\bf ij}|^2}{J_2}
+ \sum_{\bf{i}} [\frac{1}{4U} \phi_{\bf{i}}^2-\frac{i}{2}\phi_{\bf{i}}]
\nonumber \\
&-&
\frac{1}{\beta}\sum_{\bf k} \ln \{1 + \exp[-\beta(\omega_{\bf k} - \mu)]\}\ .
\label{sunfree}
\end{eqnarray}
Here the $\omega_{\bf k}$ are the eigenenergies of the mean-field
Hamiltonian $H_{MF}$: 
\begin{eqnarray}
H_{MF} = \sum_{<\bf{ij}>}[(-t_1 + \chi_{\bf{ij}}) c^{\dagger}_{\bf{i}} c_{\bf{j}} + H.c.] 
+ \sum_{<<\bf{ij}>>}[(-t_2 + \chi_{\bf{ij}}) c^{\dagger}_{\bf{i}} c_{\bf{j}} + H.c.]
+ i \sum_{\bf{i}}\phi_{\bf{i}}c^{\dagger}_{\bf{i}}c_{\bf{i}}.
\end{eqnarray}
In the zero-temperature $\beta \rightarrow \infty$ limit the fermionic contribution to
the free energy reduces to a sum over the occupied energy eigenvalues.  
The saddle point solution is found by minimizing the free energy with respect 
to $\chi$ fields, and maximizing it with respect to the $\phi$ fields. 
We carry out the minimization numerically via the  
simplex-annealing method\cite{NRC} on lattices with up to $40 \times 40$ sites.  

After ascertaining the zero-temperature phase diagram we then study the effects 
of non-zero temperature.  As the temperature is raised, $\beta \rightarrow 0$ and
the last term in equation \ref{sunfree} approaches
$k_B \ln2$ per site reflecting the fact that each site is half-occupied.  The entropy 
then dominates the free energy, terms linear in $\chi_{\bf ij}$ in equation \ref{sunfree}
disappear, and the free energy is minimized by setting 
$\chi_{\bf ij} = 0$.  Thus as the temperature is raised, 
antiferromagnetic spin correlations are weakened and then eliminated altogether. 

\subsection{Zero-Temperature Phase Diagram of the SU(N) Hubbard-Heisenberg Model}
\begin{figure}
\epsfxsize = 7cm
\hspace{5.5cm}
\epsfbox{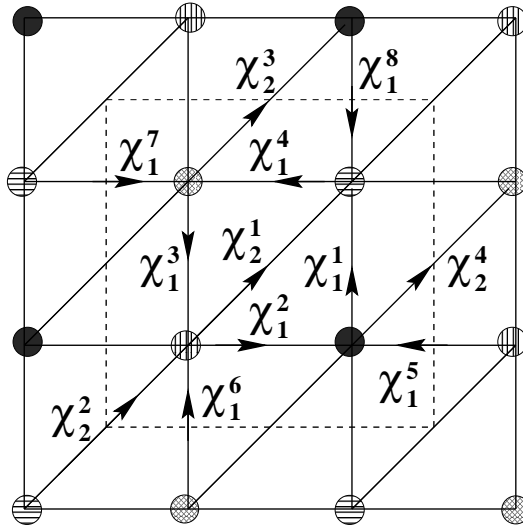}
\caption{The $2 \times 2$-site unit cell of the SU(N) Hubbard-Heisenberg model on 
the anisotropic triangular lattice.  Arrows denote the orientation of the
complex-valued $\chi_{\bf ij}$ fields.}
\label{4sitesun}
\end{figure}

We again choose a $2 \times 2$ unit cell, in anticipation that translational
symmetry can be broken at the saddle points.
The $2 \times 2$ unit cell requires 12 different complex 
$\chi$-fields and 4 different real $\phi$-fields as shown 
in figure \ref{4sitesun}.
At half-filling all $\phi_{\bf i} = 0$.  As expected, there is no site-centered 
charge density wave, and the phase diagram does not depend on the size 
of the Hubbard interaction $U$ (so long as it is repulsive) because fluctuations in 
the on-site occupancy, which are $O(\sqrt{N}) \ll N/2$  are suppressed in the 
large-N limit.  Therefore the saddle-point solutions may be classified solely 
in terms of the remaining order parameter, the $\chi$-fields.  
For the special case $t_1 = t_2 = 0$ it is important to 
classify the phases in a gauge-invariant way because there are many 
gauge-equivalent saddle-points.  In this limit there are two important 
gauge-invariant quantities: 

\noindent (i) The magnitude $|\chi_{\bf ij}|^2$ which is proportional to the 
spin-spin correlation function 
$\langle \vec{S}_{\bf i} \cdot \vec{S}_{\bf j} \rangle$.
Modulations in $|\chi|$ signal the presence of a bond-centered dimerization.

\noindent (ii) The plaquette operator $\Pi \equiv \chi_{12} \chi_{23} \chi_{34}
\chi_{41}$, where  1, 2, 3, and 4 are sites on the corners of a unit
plaquette.  By identifying the phase of $\chi$ as a spatial gauge field
it is clear that the plaquette operator is gauge-invariant, and its phase 
measures the amount of magnetic flux penetrating the plaquette\cite{marston}. 
Different saddle points are therefore gauge equivalent if the plaquette 
operator has the same expectation value, even though the $\chi$-fields may
be different.  In the Heisenberg limit, the flux always equals to $0$ or $\pi$ 
(mod $2 \pi$), so a gauge can always be found such that all the $\chi$-fields 
are purely real.  

\begin{figure}
\epsfxsize = 12cm
\hspace{2cm}
\epsfbox{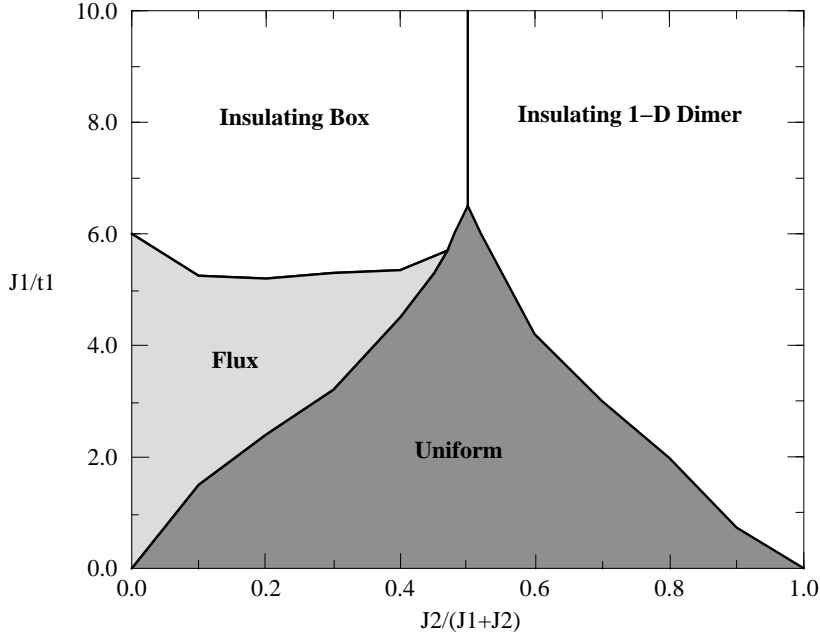}
\caption{The zero temperature phase diagram of the SU(N) Hubbard-Heisenberg model 
on the anisotropic triangular lattice. The metallic uniform region is indicated by 
dark shading while light shading demarks the semi-metallic staggered flux phase.}
\label{sunphase}
\end{figure}

Away from the pure Heisenberg limit $t_1 = t_2 = 0$ we further classify the saddle-point 
solutions in terms of whether or not they break time-reversal symmetry ($\hat{T}$). 
Finally, as there are four independent parameters ($t_1$, $t_2$, $J_1$, and
$J_2$) the phase diagram lives in a three-dimensional space of their 
dimensionless ratios.  To reduce this to a more manageable two-dimensional section, 
we assume that $J_1/J_2 = (t_1/t_2)^2$; then by varying $J_1$ and the two hopping
matrix elements we explore a two-dimensional space.  
The resulting phase diagram is shown in figure \ref{sunphase}.
We summarize the phases which appear in the diagram in the following subsections.

\subsubsection{One-Dimensional Dimer Phase}
\begin{figure}
\epsfxsize = 6.0cm
\hspace{6cm}
\epsfbox{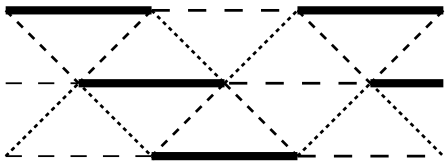}
\caption{One-dimensional dimer phase.  Dark lines indicate bonds with
strong spin-spin correlations.  These are the dimers.}
\label{dimer1d}
\end{figure}

This phase exists in the region $J_2 > J_1 > t_1$ and it exhibits spin-Peierls 
(= dimer) order.  All $\chi_2$-fields are negative real numbers with
$\chi_2^1 = \chi_2^3$, $\chi_2^2 = \chi_2^4$, and $|\chi_2^1| > |\chi_2^2|$.
Also, all $\chi_1$-fields are either small negative real numbers or
zero with $\chi_1^1 = \chi_1^3 = \chi_1^5 = \chi_1^7 < 0$,
$\chi_1^2 = \chi_1^4 = \chi_1^6 = \chi_1^8 = 0$ and 
$|\chi_1^1| \ll |\chi_2^2|$. See figure \ref{dimer1d} for a sketch.
The system breaks up into nearly decoupled dimerized spin chains. 
This phase breaks preserves $\hat{T}$-symmetry 
as all the $\chi$-fields are real. It is insulating as there is a large gap 
in the energy spectrum at the Fermi energy. 
The phase is very similar to the decoupled chain phase of the insulating Sp(N) model;
in fact the dimerization pattern is identical to one of two such possible patterns in the
Sp(N) model (see figure \ref{spnphase}) and echos the pattern found by White and 
Affleck for the two-chain zigzag model\cite{white}.  In the extreme one-dimensional limit
of $J_1 = t_1 = 0$ the SU(N) solution, like the Sp(N) solution, fails\cite{marston} to accurately
described the physics of decoupled chains, for at half-filling, the physical one-dimensional 
SU(2) Hubbard model is neither dimerized nor spin-gapped.  The inclusion of the 
biquadratic interaction suppresses dimerization\cite{matt} and yields a state qualitatively
similar to the exact solution of the physical system, but for simplicity we do not
consider such a term here.

\subsubsection{Box (also called ``Plaquette'') Phase}
\begin{figure}
\epsfxsize = 4.5cm
\hspace{6cm}
\epsfbox{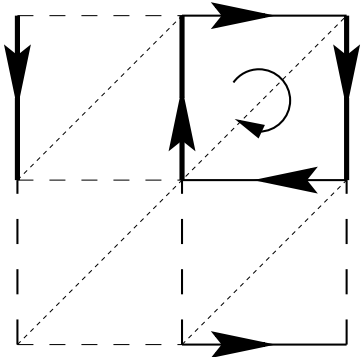}
\caption{Box (or ``plaquette'') phase with circulating currents.}
\label{box}
\end{figure}

This phase is also insulating and consists of isolated plaquettes with 
enhanced spin-spin correlations\cite{dombre,cms}.  See figure \ref{box} for a
sketch.  All $\chi_1$-fields are complex with 
$|\chi_1^1| = |\chi_1^3| > |\chi_1^2| = |\chi_1^4| >
|\chi_1^6| = |\chi_1^8| > |\chi_1^5| =|\chi_1^7|$. The $\chi_2$ fields
are small, $|\chi_2^1| = |\chi_2^2| = |\chi_2^3| = |\chi_2^4| \ll |\chi_1^5|$.
The phase $\theta$ of the plaquette product 
$\chi_1^1 \chi_1^2 \chi_1^3 \chi_1^4$ is neither 0 nor $\pi$ in general. 
The box phase breaks $\hat{T}$-symmetry when $t_1 \neq 0$.  As
time-reversal symmetry is broken, there are real orbital currents circulating
around the plaquettes\cite{ted} as shown in the figure.   Apart from 
$\hat{T}$-breaking, this phase is rather similar to the $(\pi, \pi)$ SRO phase of the Sp(N)  
model as it is a commensurate SRO phase with a large spin gap.

\subsubsection{Staggered Flux Phase (SFP)} 
\begin{figure}
\epsfxsize = 4.5cm
\hspace{6cm}
\epsfbox{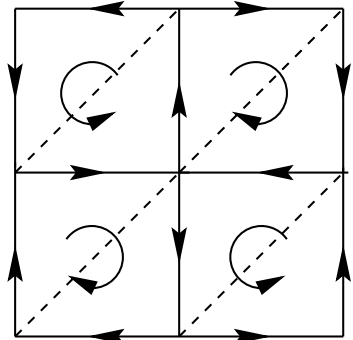}
\caption{Staggered flux phase (SFP) with circulating currents.}
\label{flux}
\end{figure}

All $\chi_1$-fields are equal, with an imaginary component, and
the $\chi_2$-fields are equal, real, and much smaller in magnitude than
the $\chi_1$ fields.  The phase of the plaquette operator 
differs in general from 0 or $\pi$; see figure \ref{flux} for a sketch.
The staggered flux phase breaks $\hat{T}$-symmetry.
Like the box phase, there are real orbital currents circulating around the 
plaquettes\cite{ted}, in an alternating antiferromagnetic pattern.  
The SFP is semi-metallic as the density of states (DOS) is small, and in fact
vanishes linearly at the Fermi energy at $J_2 = 0$.  Apart from $\hat{T}$-breaking
this phase is rather similar to the $(\pi, \pi)$ LRO phase of the Sp(N) antiferromagnet,
as the spins show quasi-long-range order with power-law decay in the spin-spin 
correlation function.  In the limiting case of a pure Heisenberg AF on the nearest-neighbor 
square lattice, $t_1 = t_2 = J_2 = 0$, gauge fluctuations at sufficiently small-N are expected to
drive the SFP (which can be stabilized against the box phase by the addition of the 
biquadratic interaction) into a $(\pi, \pi)$ N\'eel-ordered state\cite{gauge}.  

\subsubsection{Uniform Phase} 
\begin{figure}
\epsfxsize = 4.5cm
\hspace{6cm}
\epsfbox{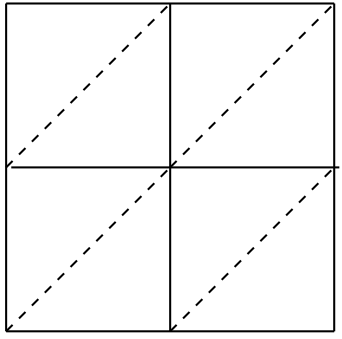}
\caption{Uniform phase with no broken symmetries.  The uniform phase is a Fermi liquid.}
\label{uniform}
\end{figure}

All $\chi$-fields are negative real numbers and
$\chi_1^1 = \chi_1^2 = \cdots = \chi_1^8$,
$\chi_2^1 = \chi_2^2 = \chi_2^3 = \chi_2^4$ with $\chi_1^1 \neq \chi_2^1$
in general. See figure \ref{uniform} for
a sketch.  This phase preserves $\hat{T}$-symmetry and 
since all $\chi$-fields are real, they 
simply renormalize hopping parameters $t_1$, $t_2$.  The uniform phase is therefore 
a metallic Fermi liquid.  Spin-spin correlations 
in the uniform phase decay as an inverse power law of the separation, 
with an incommensurate wavevector, so the uniform phase behaves  
similarly to that in the $(q, q)$ LRO phase of the Sp(N) model (see below).  

\subsection{Global Similarities Between SU(N) and Sp(N) Phase Diagrams}
There are some global similarities between the SU(N) phase diagram of the 
Hubbard-Heisenberg model (figure \ref{sunphase}) and the Sp(N) phase diagram of the
insulating Heisenberg antiferromagnet (figure \ref{spnphase}).  We have already pointed out 
similarities between the four phases of the SU(N) model and the decoupled chain, 
$(\pi, \pi)$ SRO, $(\pi, \pi)$ LRO, and $(q, q)$ LRO phases of the Sp(N) model.  
Apparently the dimensionless
parameter $J_1/t_1$ on the vertical axis of the SU(N) phase diagram (figure \ref{sunphase})
is the analog of the quantum parameter $1/\kappa$ of the Sp(N) phase diagram.  The
reason why these two dimensionless parameter play similar roles can be understood
by considering the limit of the pure insulating antiferromagnet, corresponding to 
$t_1 \rightarrow 0$ and $t_2 \rightarrow 0$.  In these limits, the SU(N) model can only
be in the purely insulating box phase, or in the dimerized phase.  The
spins are always quantum disordered, and behave like the extreme 
quantum $\kappa \rightarrow 0$ limit of the Sp(N) model.  In the opposite limit 
$t_1 \rightarrow \infty$ and $t_2 \rightarrow \infty$ the spin-spin correlation
function decays more slowly, as an inverse power law instead of exponentially.  This
is as close to LRO as is possible in the large-N limit of the SU(N) model.  
Roughly then it corresponds to the ordered 
classical $\kappa \rightarrow \infty$ limit of the Sp(N) model.     

\subsection{Observable Properties of the SU(N) Hubbard-Heisenberg Model}
We now comment on the possible relevance of our large-N solution to 
the electronic and magnetic properties of the organic superconductors.
In reference \onlinecite{mckenzie} it is estimated
that $J_2/J_1 \sim 0.3$ to $1$ and $J_1/t_1 \sim 0.5$ to $2$.
Figure \ref{sunphase} then implies that the ground
state is the uniform phase, which as noted above, is a rather ordinary
Fermi liquid with no broken symmetries.  

We note that in the one-dimensional limit 
($t_1 = J_1 = 0$) and in the square lattice limit ($t_2 = J_2 = 0$), 
nesting of the Fermi surface is perfect, and the system is driven into an
insulating phase no matter how large the hopping amplitudes.  Away from these 
two extreme limits, however, there is a non-trivial metal-insulator 
transition line.  This accords with expectations because as
the Fermi surface is not perfectly nested, the metallic state is only 
eliminated at a nonzero value of $J_1/t_1$.  Experimentally it is 
found that upon increasing pressure, a SIT transition from the 
antiferromagnetic insulator to a superconductor is seen in the 
$\kappa$-(BEDT-TTF)$_2$X family of materials\cite{kanoda,kino}.  
This transition can be understood in terms of the SU(N) phase diagram as follows.  
As pressure increases, the effective hopping amplitudes $t_1$ and $t_2$ also increase 
because Coulomb blocking is less effective\cite{mckenzie}.  The bandwidth
broadens, but the spin exchange couplings $J_1$ and $J_2$ remain nearly unchanged 
as these are determined by the {\it bare}, not the effective, hopping matrix elements. 
Thus the ratio $J_1/t_1$ is reduced at high pressure, and the many-body correlations 
weaken.  From the phase diagram (figure \ref{sunphase}) it is apparent that the system 
can be driven from an insulating state into a metallic state, 
which presumably superconducts at sufficiently
low temperature.  Thus the transition to a conducting state at high pressure 
can be seen as being due to bandwidth broadening, 
as in the Brinkman-Rice picture of the MIT.

\begin{figure}
\epsfxsize = 11cm
\hspace{2.5cm}
\epsfbox{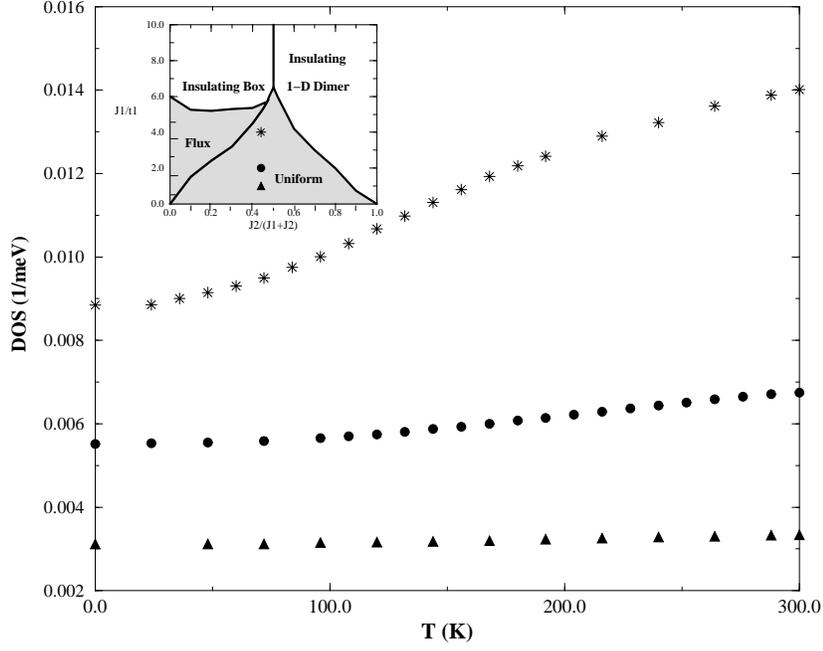}
\caption{Density of states per unit volume at the Fermi energy
as a function of temperature for three points, 
indicated by the inset and specified in table \ref{table1}, 
inside the uniform phase.}
\label{DOS}
\end{figure}

\begin{table}
\caption{Parameters (in meV) for the three points in figure \ref{DOS}.} 
\begin{tabular}{|c|c|c|c|c|c|c|}
Symbol In Figure \ref{DOS} & $t_1$ & $t_2$
 & $J_1$ & $J_2$  & $J_1 / t_1$ & $J_2 / (J_1 + J_2)$  \\
\tableline
star & 6 & 5.4 & 24 & 19.2 & 4 & 0.44 \\
dot & 12 & 10.8 & 24 & 19.2& 2 & 0.44  \\
triangle  & 24 & 21.6 & 24 & 19.2 & 1 & 0.44 \\
\end{tabular}
\label{table1}
\end{table}

We plot the DOS as a function of temperature in figure \ref{DOS} for three
points inside the uniform phase identified in table \ref{table1}.
The DOS decreases as the temperature decreases from room temperature
down to absolute zero.  Clearly the pseudogap is more prominent 
near the boundary between the conducting uniform phase and the semimetallic SFP and 
insulating box phases.  The drop in the DOS could qualitatively 
explain the $30\%$ depression seen in the uniform susceptibility\cite{kanoda} 
(NMR experiments find a reduction of about $50\%$ as the temperature decreases
from $100~ ^\circ$K to $10~ ^\circ$K\cite{desoto,mayaffre}).
The behavior may be understood
as follows.  As the temperature decreases, antiferromagnetic spin-spin 
correlations develop.  These correlations are signaled in the mean-field theory by 
the link variables $\chi_{\bf ij}$ which become non-zero at low enough temperature.  
Electrons on neighboring sites then tend to have 
opposite spin, reducing Pauli-blocking, and making their 
kinetic energy more negative.  The band widens and the density of states drops. 

\section{Discussion}
\label{sec:discuss}

We have solved the bosonic Sp(N) Heisenberg and 
fermionic SU(N) Hubbard-Heisenberg models on
the anisotropic triangular lattice in the large-N limit.  The bosonic Sp(N) 
representation of the Heisenberg model is useful for describing magnetic 
ordering transitions.  It therefore may be an appropriate description of
the insulating phase of the layered organic superconductors 
$\kappa$-(BEDT-TTF)$_2$X and of the insulating Cs$_2$CuCl$_4$ compound.  
The fermionic SU(N) Hubbard-Heisenberg model
provides a complementary description of the charge sector, in particular 
the physics of metal-insulator transitions and the
unconventional metallic phases.  Systematic expansions in powers of $1/N$ 
about the large-N limit are possible for either model.

For the Sp(N) model we found five phases: (i) three commensurate phases, the  
$(\pi, \pi)$ LRO and SRO phases, and the decoupled chain phase and (ii) two
incommensurate phases, the $(q, q)$ LRO and SRO phases.  Passing from the square 
lattice ($J_2 = 0$) to the one-dimensional limit ($J_1 = 0$) at
$\kappa = 1$ which corresponds to spin-1/2 in the physical Sp(1) limit
first there is the N\'eel ordered phase, 
the incommensurate $(q, q)$ LRO phase, the incommensurate $(q, q)$ SRO 
phase, and finally a phase consisting of decoupled chains. 
These phases are similar to those obtained from a series expansion 
method\cite{zhang} and a weak coupling renormalization group 
technique\cite{tsai}.  The effects of finite-N fluctuations on the saddle-point 
solutions were also discussed.  The observed dispersion of spin excitations 
in the Cs$_2$CuCl$_4$ material\cite{coldea} can be described either as
spinwaves in the incommensurate $(q, q)$ LRO phase, or in terms of 
pairs of spinons in the deconfined $(q, q)$ SRO phase.

The zero-temperature phase diagram of the SU(N) Hubbard-Heisenberg model in
the large-N limit has a 1-D dimer phase, a box (or plaquette) phase, a staggered-flux 
phase, and a uniform phase.  In the extreme 1-D and square lattice limits, 
the ground state of the half-filled Hubbard-Heisenberg
model is always insulating because the nesting of the Fermi surface is perfect.
But away from these two extreme limits there is a metal-insulator transition.  
For parameters appropriate to the $\kappa$-(BEDT-TTF)$_2$X class of materials 
we find that the conducting regime is described by our uniform phase, which is
a rather conventional Fermi liquid with no broken symmetries.  In this phase 
we find that the density of states at Fermi level decreases at low temperatures,
due to the development of antiferromagnetic correlations. 
This could explain the depression seen in the uniform susceptibility 
of the organic superconducting materials at low temperatures.  
It agrees qualitatively with experiments which suggest the existence of
a pseudo-gap. 

We have used two models instead of one because the two large-N theories have 
complementary advantages and disadvantages.  The bosonic Sp(N) approach is 
not suitable for describing electronic properties, as there are no 
fermions.  The fermionic SU(N) approach, on the other hand, is not useful 
for describing magnetic ordering, as the order parameters are all SU(N)
singlets.  Neither mean-field approximation can describe the superconducting phase.  
The {\it fermionic} Sp(N) Hubbard-Heisenberg model does support superconducting 
states, but no non-superconducting metallic phases\cite{matthias}.
Whether or not a single model can be constructed which is exact in a 
large-N limit and yet encompasses all the relevant phases remains an open
problem.  

\acknowledgements

We thank Jaime Merino, Bruce Normand, Nick Read, Shan-Wen Tsai, and Matthias Vojta 
for helpful discussions.  We especially thank Subir Sachdev for his insightful comments 
about the Sp(N) phase diagram and the nature of the deconfined phases.  
This work was supported in part by the NSF under grant Nos. DMR-9712391
and PHY99-07949.  J.B.M. thanks the Institute for Theoretical Physics at UCSB for
its hospitality while he attended the ITP Program on High Temperature Superconductivity.
Computational work in support of this research was carried out using
double precision C++ at the Theoretical Physics Computing Facility 
at Brown University.  Work at the University of Queensland was supported by
the Australian Research Council.

\end{document}